%% file: FFP14.tex
\documentclass{PoS}
\usepackage{fontenc,indentfirst,delarray,amsmath,amssymb}
\usepackage{graphicx,graphics,epstopdf,epsf}
\usepackage{pdfsync}
\def\OO{{\mathcal O}}
\input{commandes}

\title{Designing the sound of a cut-off drum}
\ShortTitle{Cut-off drum}

\author{\speaker{Pierre Martinetti}\\\thanks{Supported by the Italian
    project ``Prin 2010-11 - Operator Algebras, Noncommutative
    Geometry and Applications''.}\\
        Dipartimento di Matematica, Universit\`a di Trieste\\
        E-mail: \email{pmartinetti@units.it}}

\abstract{The spectral action in noncommutative geometry naturally
  implements an ultraviolet cut-off, by counting the eigenvalues
  of a (generalized) Dirac operator lower than an energy of
  unification. Inverting the well known question ``how to hear the shape
  of a drum'', we ask what drum can be designed by hearing the
  truncated music of the spectral action ? This makes sense because
  the same Dirac operator also determines the metric, via Connes
  distance. The latter thus offers an original way to implement the
  high-momentum cut-off of the spectral action as a short distance
  cut-off on space. This is a non-technical presentation of the results of
  \cite{DAndrea:2013kx}.}

\FullConference{Frontiers of Fundamental Physics 14 - FFP14,\\
                15-18 July 2014\\
                Aix Marseille University (AMU) Saint-Charles Campus, Marseille }
		
\begin{document}

\section{Introduction}

Cut-off are  generally used to avoid undesirable divergencies
occurring at small or very large scales. The scale is
usually an energy scale $E$ and the
cut-off  is implemented either on  momentum $p$ or on the wavelength
$\lambda$.  In various senses, these two implementations are dual to each
other:  a high momentum cut-off is equivalent to  a short wavelength
cut-off and vice-versa, as can be read e.g. in de Broglie
relation $p = \frac h{\lambda}$. In this note we explore another
possibility to give sense to this duality,
based on the double nature of the
(generalized) Dirac
operator $D$ which is at the heart of Connes approach to
noncommutative geometry \cite{Connes:1994kx}. This is indeed the same
operator $D$ which

- provides an action which naturally
incorporates a high energy cut-off. This is the 
\emph{spectral action} \cite{Chamseddine:1996kx}
\begin{equation}
  \label{eq:26}
  \text{Tr}\; f\left(\frac{D}{\Lambda}\right)
\end{equation}
where $f$ is the characteristic function of the interval $[0,1]$ and
$\Lambda$ a energy scale of unification. We refer to
\cite{Chamseddine:2007oz} for details on the choice of
the operator $D$ and how the
asymptotic expansion $\Lambda\to\infty$ yields the standard model of elementary particles minimally
coupled with (Euclidean) general relativity (see also 
\cite{Walterlivre} for an highly readable introduction to the subject,
 \cite{Chamseddine:2013uq, Chamseddine:2013fk} and
\cite{Devastato:2013fk,buckley} for recent developments).

- defines a metric on the space ${\cal S}(\A)$ of states{\footnote{{A state $\varphi$ of $\A$  is a linear application $\A\to\mathbb C$
which is positive ($\varphi(a^*a)\in \R^+$) and of norm $1$ (in case
$\A$ is unital, this means $\varphi(\mathbb I)=1$).}} of an algebra
$\A$, provided the later acts on the same Hilbert space $\cal
H$ as $D$ in such a way that 
\begin{equation}
 L_D(a) :=||[D,\pi(a)]||\label{eq:5}
 \end{equation}
is finite for any $a\in\A$ ($\pi$ denotes the representation of $\A$ as
bounded operators on $\cal H$). This is the \emph{spectral
  distance} \cite{Connes:1989fk}
\begin{equation}
d_{\A, D} (\varphi,\psi) := \underset{a\in \A}{\sup}\{
\abs{\varphi(a) - \psi(a)}, \,  L_D(a)\leq 1\}
\label{eq:4}
\end{equation} 
for any  $\varphi, \psi$ in ${\cal S}({\A})$.
For $\A= \cinf$ the algebra of smooth functions on a manifold $\M$ and $D=\slash\!\!\!\!\!\partial$ the usual
Dirac operator of quantum field theory, the spectral distance computed
between pure states (which are nothing but the points of $\M$, viewed
as the application $\delta_x: f\to f(x)$) gives back
the geodesic distance \cite{Connes:1992bc},
\begin{equation}
d_{\cinf, \slash\!\!\!\!\!\partial}(\delta_x,
  \delta_y) = d_{\text{geo}}(x, y).\label{eq:32}
  \end{equation}

The action \eqref{eq:26}
counts the eigenvalues of the Dirac operator smaller than
the energy scale $\Lambda$, which amounts to cut-off the Fourier modes
with energy greater than $\Lambda$. In other terms the spectral action
naturally implements an ultraviolet cut-off as a high momentum
cut-off. The question we adress in this note is: can this be read as a short
wave-length cutoff in the distance formula~\eqref{eq:4}~?  More
precisely, by
cuting-off the spectrum of the Dirac operator
in the distance formula, does one transform the high-momentum cut-off
$\Lambda$ in a short distance cut-off $\lambda$ ? Reversing the
well known question on how to retrieve the shape of a drum
from its vibration modes, our point here is to understand what drum can one
design from hearing the spectral action ?
We  have shown in \cite{DAndrea:2013kx} that the answer is not obvious, and
asks for a careful discussion on the nature of the points
of the ``cut-off drum''. We report here some of these results, in a non
technical manner.

\section{Cutting-off the geometry}
 
We implement a cut-off by the conjugate action
of a projection $P_\Lambda$ acting on $\cal H$: 
  \begin{equation}
D\to \ {D_\Lambda:= P_\Lambda \,D \,P_\Lambda}.
\label{eq:1}
\end{equation}
Typically $P_\Lambda$ is a projection on the eigenspaces of $D$, but
for the moment it is simply a projection acting on $\cal H$. We
are interested in computing the spectral distance on a manifold $\M$
induced by this cut-off, that is formula \eqref{eq:4} for $\A=\cinf$ but
with $D$ substituted with $D_\Lambda$.

Suppose that $D_\Lambda$
is a bounded operator with norm $\Lambda\in\R^+$. Then one has
\cite[Prop. 5.1]{DAndrea:2013kx}
  \begin{equation}
 d_{\cinf, D_\Lambda}(\delta_x, \delta_y)\geq \Lambda^{-1}.
\label{eq:16}
 \end{equation}
This seems to be precisely the answer one was expecting: by cutting-off
the spectrum of $D$ below $\Lambda$, one is not able to probe space with a resolution
better than $\Lambda^{-1}$. Unfortunately \eqref{eq:16} is an
inequality, not an equality. There is
as expected a lower bound to the resolution on the position space, but
nothing guarantees that this bound is optimal. In particular if $D_\Lambda$ has finite rank, then the
distance is actually infinite \cite[Prop.5.4]{DAndrea:2013kx}. For
$\M$ compact, this happens for instance
when $P_\Lambda = P_N$ is the projection on the first
$N$ Fourier modes for some $N\in\mathbb N$. Then $D_\Lambda =  D_N := P_N D P_N$ has finite rank
and the distance between any two points is infinite,
\begin{equation}
d_{\cinf, D_N}(\delta_x, \delta_y)=\infty.
\label{eq:25}
\end{equation}
In other terms,
cutting-off all but a finite number of Fourier modes destroys the metric
  structure of the manifold. It is an open question whether the distance remains
  finite for a bounded  $D_\Lambda$ with infinite rank.
\smallskip

Eq. \eqref{eq:25} illustrates the tension between truncating the Dirac
operator while keeping the usual notion of points. A solution to maintain a metric
structure is to truncate point as well. This actually makes sense in full
generality, that is for $\A$ non-necessarily commutative, acting on some Hilbert
space $\cal H$ together with an operator $D$ such that $L_D(a)$
is finite for any $a\in\A$. Given a finite rank
projection $P_N$ in $\mathcal B(\HH)$, we then define
\begin{equation}
{\mathcal O_N := P_N \,\mathcal
\pi(\A_{sa})\, P_N }
\label{eq:9}
\end{equation}
where $\A_{sa}$ is the set of selfadjoint elements of $\A$.  The set ${\cal
  O}_N$  has no reason to be an algebra but it has the structure of
\emph{ordered unit space}, which is sufficient to define its state
space ${\cal S}({\cal O}_N)$ and to give sense to formula
\eqref{eq:4} (substituting the seminorm $L_D$ with the seminorm
\begin{equation}
  \label{eq:33}
  L_N:=||[D_N, \cdot]||
\end{equation}
and $\A$ with ${\cal O}_N$). In addition to the
original distance $d_{\A, D}$, one thus inherits from the cut-off two ``truncated'' distances:
$d_{\A, D_N}$ on ${\cal S}(\A)$ and $d_{{\cal O}_N, D_N}$ on ${\cal
  S}({\cal O}_N)$.  To make the comparison between these distances possible, 
we use the injective map
$\varphi^\sharp:=\varphi\circ\text{Ad} \,P_N$ that sends a state
$\varphi$ of
$\OO_N$ to a state $\varphi^\sharp$ of $\A$. By pull back,
one obtains
three distances on ${\cal S}({\cal O}_N)$:
\begin{equation}
d_{{\cal O}_N, D_N}(\varphi, \psi),\quad d_{\A,D}^{\,\flat}(\varphi,\psi) :=
d_{\A,D}(\varphi^\sharp,\psi^\sharp), \quad d_{\A,D_N
}^{\,\flat}(\varphi,\psi):= d_{\A,D_N
}(\varphi^\sharp,\psi^\sharp).\label{eq:11}
\end{equation}
A sufficient condition that makes the truncated distance $d^\flat_{\A, D_N}$
equivalent to  the ``bi-truncated'' distance $d_{{\cal O}_N, D_N}$
is that \cite[Prop. 3.5]{DAndrea:2013kx}  the
seminorm $L_N:=||[D_N, \cdot]||$ is \emph{Lipschtiz} \cite{rieffel2003}, meaning
that  $L_N(a)= 0$ if and only if $a={\mathbb C} {\bf 1}$.
If in addition the non-truncated seminorm $L_D$ is Lipschitz, or $P_N$
is in the commutant $\A'$ of the algebra $\A$, or $P_N$ commutes with
$D$, then one also has that $d^\flat_{\A, D}$ is equivalent to
$d_{{\cal O}_N, D_N}$.

In the commutative case, the set ${\cal S}({\cal O}_N)$ permits to give
 a precise meaning to the notion of \emph{truncated points}. By this
 we mean a sequence of states of
 ${\mathcal O}_N$ that tends to some $\delta_x$ 
 as $N\to\infty$. For
instance on the circle, that is $\A = C^\infty(S^1)$,  the Fejer transform 
\begin{equation}
\Psi_{x,N}(f) = \sum_{n=-N}^{N}(1-\frac{|n|}{N+1})f_ne^{inx}, \quad
N\in\mathbb N
\label{eq:8}
\end{equation}
is a state of ${\mathcal O}_N$ for $P_N$ the projection on the first
$N$ negative and $N$ positive Fourier modes
\cite[Lem.~5.10]{DAndrea:2013kx}. It is an approximation of the point $x\in S^1$ in that
\begin{equation}
  \label{eq:6}
  \text{lim}_{N\to\infty} \Psi_{x,N}(f) = f(x) \quad \forall f\in C^\infty(S^1).
\end{equation}
Moreover this approximation deforms the metric structure of the circle
but does not destroy it, since  - with $D$ the
usual Dirac operator of $S^1$ - the bi-truncated distance between any two Fejer
transforms is finite for any $N$  \cite[Prop.~5.11]{DAndrea:2013kx},
  \begin{equation}
d_{{\cal O}_N, D_N}(\Psi_{x,N},\Psi_{y,N}) \leq d_{\mathrm{geo}}(x,y),
\label{eq:29}
\end{equation}
and tends towards the geodesic distance  for large $N$,
\begin{equation}
\lim_{N\to\infty} d_{{\cal O}_N, D_N}(\Psi_{x,N},
\Psi_{y,N})=d_{\mathrm{geo}}(x,y)\quad \quad\forall x,y\in
S^1.
\label{eq:30}
\end{equation}
A similar example has been worked out on the real line \cite[Prop.~5.7]{DAndrea:2013kx}.

\section{Convergence of truncations}

Let us study in a more systematic way the idea introduced in the
previous section of approximating a state by a sequence of truncated
states. To this aim, take $\A$, $\cal H$ and $D$ satisfying the
conditions of the precedent section, and let us consider a sequence $\left\{P_N\right\}_{N\in\mathbb N}$ of
increasing finite-rank projections, weakly converging to~$1$. Under
which conditions can states of $\A$ be approximated by states of
${\mathcal O}_N$ in such a way that the metric structure is preserved
? 

For normal states {\footnote{$\varphi\in\sa$ is normal if
    and only if there exists a positive traceclass
operator $R$ on $\cal H$ (the density matrix) such that $\varphi(a) =
\text{Tr}(Ra),\; \forall a\in \A$.}}, the answer is simple in case the topology induced by the spectral distance coincides with the
weak$^*$ topology.  Then any normal states $\varphi$ with density
matrix $R$ is the limit of its
truncation \cite[Prop. 4.2]{DAndrea:2013kx}, 
\begin{equation}
  \label{eq:27}
\lim_{N\to\infty}  d_{\A,D}(\varphi, \varphi_N^\sharp) = 0
\end{equation}
where $\varphi_N$ is the state with density matrix $Z_N^{-1}R$ where $Z_N:=\text{Tr}(P_NR)$.
One also has the convergence in the sense of metric spaces\cite[Prop. 4.3]{DAndrea:2013kx}:
$({\cal S}(\OO_N),d^\flat_{\A,D})$
converges to $(\,\overline{{\cal N}(\A)},d_{\A,D})$ for the
Gromov-Hausdorff distance.
\medskip

In case the topology of the spectral distance is not the weak$^*$
topology, the answer is more challenging. A preliminary step is to
work out a class of states at finite distance from one another. In
the commutative case, for $\M$ connected and complete, such a class is given by states with finite moment of
order $1$. Recall that there is a $1$-to-$1$ correspondance
between states $\varphi$ of $C_0^\infty(\M)$ and probability measures
$\mu$ on $\M$,
\begin{equation*}
  \varphi(f) = \int_\M f(x)\;  \de\mu(x) \quad \forall f\in C_0(\M).
\end{equation*}
For $\M$ connected, the finiteness of the moment of order $1$ of $\varphi$, 
\begin{equation}
  {\mathcal M}_1(\varphi, x') := \int_\M d_{\text{geo}}(x, x') \,\de\mu(x) \;
\end{equation}
does not depend on
the choice of $x'\in\M$ and so is intrinsic to the state. If
furthermore $\M$ is complete, one has
that the spectral distance $d_{\slash\!\!\!\!\!\partial}$ between states with finite
  moment of order $1$ is finite (see e.g. \cite{dAndrea:2009xr}).

In the noncommutative case, the correspondance between states and
probability measure on the pure state space is no longer $1$-$1$, as
can be seen on easy examples such as $M_2(\mathbb C)$. However
in \cite[\S 4.2] {DAndrea:2013kx} we proposed to give meaning to the notion of ``finite moment of order
$1$'' for normal states in the following way.  Let 
$\varphi$ be a normal state with density matrix $R$. Fix an 
orthonormal basis $\mathfrak{B}=\{\psi_n\}_{n\in\N}$ of
$\HH$ made of eigenvectors of $R$, with eigenvalues $p_n\in\R^+$.
 Denote $\Psi_n(a):=\inner{\psi_n,a\psi_n}$  the corresponding vector
 states in ${\cal S}(\A)$ so that
\begin{equation*}
\varphi(a) = \sum_{n\geq 0} p_n \, \Psi_n(a) \quad \forall a\in\A .
\end{equation*}
  We call  \emph{moment of order $1$ of $R$} with respect to  the eigenbasis $\mathfrak{B}$
  and to  a state $\Psi_k$ (induced by a vector $\psi_k\in
  \mathfrak{B}$) the quantity
\begin{equation}
    \label{eq:7}
    {\mathcal M}_1(R,\mathfrak{B},\Psi_k):= \sum_{n\geq 0} p_n \, d_{\A, D}(\Psi_k, \Psi_n).
  \end{equation}
Unlike the commutative case the finiteness of \eqref{eq:7}
is not intrinsic to the density matrix (hence even less to
the state), because for the
same density matrix $R$ one may have that 
${\mathcal M}_1(R,\mathfrak{B},\Psi_k)$ is infinite for a given basis while ${\mathcal
  M}_1(R,\mathfrak{B}',\Psi'_k)$ is finite for another one \cite[Ex. 4.6] {DAndrea:2013kx}. However, 
once fixed $\mathfrak{B}$,
the finiteness of ${\mathcal M}_1(R,\mathfrak{B}, \Psi_k)$ does not
depend on $\Psi_k$. 
We write $\NN_0(\A)$ the set of normal states for which there exists at least one density matrix $R$ with an eigenbasis $\mathfrak{B}=\{\psi_n\}$
such that
\begin{equation}
{\mathcal M}_1(R,\mathfrak{B}, \Psi_n)<\infty.
\label{eq:10}
\end{equation}

Consider then an increasing sequence $\left\{P_N\right\}_{N\in\N}$ of
projections weakly convergent to $1$. 
For any $\varphi\in\NN_0(\A)$ such that \eqref{eq:10} holds for an eigenbasis $\mathfrak{B}$
  in which the $P_N$'s are all diagonal,
there exists a sequence $\varphi_N\in {\cal S}(\OO_N)$ such that
\begin{equation}
\underset{N\to\infty}{\lim} d_{\A, D}(\varphi, \varphi^\sharp_N) =0.
\end{equation}
In other terms, any $\varphi\in\NN_0(\A)$ can be approximated in the metric
  topology by a truncation $\varphi_N$. However unlike \eqref{eq:27}
  where the
  truncating-projections $P_N$ where fixed once for all, in case the
  metric topology is not the weak$^*$ the truncating procedure may
  depend on the state.

\section{An unbounded example: Berezin quantization of the plane}

We conclude by an example where the truncated Dirac operator
is not bounded: the Berezin quantization of the plane. We omit the
details that can be found in \cite[\S 6.2]{DAndrea:2013kx}. For other
applications of noncommutative geometry to Berezin quantization, see \cite{Englis:2014aa}.


One starts with $\HH =L^2(\C, \frac{d^2z}{\pi})$ and, for   $\theta>0$, 
define $P_\theta$ as the projection on the subspace
\begin{equation}
\HH_\theta := \text{Span}\left\{
 h_n(z):=
 \frac{z^n}{\sqrt{\theta^{n+1}n!}}\,e^{-\frac{|z|^2}{2\theta}}
\right\}_{n\in\N}.
\label{eq:14}
\end{equation}
For $D$ the Dirac operator of the Euclidean
plane, the truncated Dirac operator
\begin{equation}
D_\theta:=(P_\theta\otimes\I_2) \,D\, (P_\theta\otimes\I_2) = \frac{2}{\sqrt\theta}
\left(
\begin{array}{cc}
 0 & {\mathfrak a}^\dag\! \\ 
{\mathfrak a} & 0\, 
\end{array}
\right)\label{eq:21}
\end{equation}
is unbounded (${\mathfrak a}^\dag, {\mathfrak a}$ are the creation, annihilation operators
on the $h_n$'s).

Let $\A={\cal S}(\R^2)$ denote the algebra of Schwartz functions on the
plane,  and denote $\OO_\theta$ the order unit space generated by $P_\theta \,f \,P_\theta$ (for $f=f^*\in\A$).
Both  act on $\HH\otimes\C^2$. For any states $\varphi, \psi$ of $\A$, define
\begin{equation}
d_{\A,D}^{(\theta)}(\varphi,\psi)
:=\sup_{f=f^*\in\A}\left\{\varphi(f) - \psi(f),\, ||[D,
    B^\theta(f)||\leq 1\right\}\label{eq:24}
  \end{equation}
where 
\begin{equation}
  \label{eq:34}
  B_\theta(f): z\to \scl{\psi_z}{P_\theta f P_\theta\,\psi_z} \quad\text{ where }\quad
\psi_z=e^{-\frac{|z|^2}{2\theta}}\sum_{n=0}^\infty \frac{\bar
    z^n}{\sqrt{\theta^n n!}}h_n
\end{equation}
 is the \emph{Berezin transform} of $f$. One gets
\cite{DAndrea:2013kx}
\begin{equation}
d_{\A,D}(\varphi^\sharp,\psi^\sharp)\leq
d_{\OO_\theta,D_\theta}(\varphi,\psi)\leq d_{\A,D}^{(\theta)}(\varphi^{\sharp},\psi^{\sharp}).
\label{eq:18}
\end{equation}
In particular, the distance between coherent states $\Psi_z, \Psi_{z'}$, $z, z'\in\C$, is 
\begin{equation}
d_{\OO_\theta, D_\theta}(\Psi_z, \Psi_{z'}) = |z -z'|.
\label{eq:19}
\end{equation}
A similar result was found in \cite{Martinetti:2011fko} from a
completely different perspective, based on the construction of the
element that attains the supremum in the distance formula. This
illustrates that the cut-off procedure could be an efficient tool to
make explicit calculations of the distance.
 
\bibliographystyle{abbrv}
\bibliography{/Users/pierre/physique/articles/Bibdesk/biblio}
\end{document}

%% file: commandes.tex
\definecolor{bleuvert}{rgb}{.1,.5,.4}
\definecolor{light-gray}{gray}{0.95}
\definecolor{gray}{gray}{0.75}
\definecolor{violet}{rgb}{0.4,0.,0.3}
\definecolor{jaune}{rgb}{0.8,0.6,0.1}
\definecolor{cvert}{rgb}{0.8,0.6,0.5}

\newcommand{\abs}[1]{\lvert#1\rvert}

\newcommand{\scl}[2]{\langle#1,#2\rangle}

\def\begf{\begin{frame}}
\def\enf{\end{frame}}

\def\begz{\begin{itemize}}

\def\endz{\end{itemize}}

\def\lp{\left(} 
\def\rp{\right)} 
\def\dm{\lp\begin{array}}	
\def\fm{\end{array}\rp}
\def\m2{M_2 \lp \cc \rp}
\def\m3{M_3 \lp \cc \rp}

\def\cc{{\mathbb{C}}}
\def\C{{\mathbb{C}}}	

\def\R{{\mathbb{R}}}

\def\N{{\mathbb{N}}}

\def\I{{\mathbb{I}}}

\def\mm{{\mathcal M}}	
\def\M{{\mathcal M}}		
\def\aa{{\mathcal A}}

\def\A{{\mathcal A}}

\def\cinf{C^{\infty}\lp\mm\rp}

\def\xo0{\omega^0_x}
\def\yo0{\omega^0_y}

\def\xo0{x_\omega^0}
\def\yo0{y_\omega^0}

\def\sa{{\cal S}(\aa)}

\def\fm{\Phi(x^\mu)}

\def\dm{\partial_\mu}

\def\NN{{\cal N}}
\def\dmm{\left(\begin{array}}
\def\fmm{\end{array}\right)}

\newcommand{\HH}{\mathcal{H}}
\newcommand{\de}{\mathrm{d}}

\newcommand{\inner}[1]{\left<#1\right>}

%% file: FFP14.bbl
\begin{thebibliography}{10}

\bibitem{Chamseddine:1996kx}
A.~H. Chamseddine and A.~Connes.
\newblock The spectral action principle.
\newblock {\em Commun. Math. Phys.}, 186:737--750, 1996.

\bibitem{Chamseddine:2007oz}
A.~H. Chamseddine, A.~Connes, and M.~Marcolli.
\newblock Gravity and the standard model with neutrino mixing.
\newblock {\em Adv. Theor. Math. Phys.}, 11:991--1089, 2007.

\bibitem{Chamseddine:2013uq}
A.~H. Chamseddine, A.~Connes, and W.~van Suijlekom.
\newblock Beyond the spectral standard model: emergence of {P}ati-{S}alam
  unification.
\newblock {\em JHEP}, 11:132, 2013.

\bibitem{Chamseddine:2013fk}
A.~H. Chamseddine, A.~Connes, and W.~van Suijlekom.
\newblock {I}nner fluctuations in noncommutative geometry without first order
  condition.
\newblock {\em J. Geom. Phy.}, 73:222--234, 2013.

\bibitem{Connes:1989fk}
A.~Connes.
\newblock Compact metric spaces, {F}redholm modules, and hyperfiniteness.
\newblock {\em Ergod. Th. \& Dynam. Sys.}, 9:207--220, 1989.

\bibitem{Connes:1994kx}
A.~Connes.
\newblock {\em Noncommutative Geometry}.
\newblock Academic Press, 1994.

\bibitem{Connes:1992bc}
A.~Connes and J.~Lott.
\newblock The metric aspect of noncommutative geometry.
\newblock {\em Nato ASI series B Physics}, 295:53--93, 1992.

\bibitem{DAndrea:2013kx}
F.~D'Andrea, F.~Lizzi, and P.~Martinetti.
\newblock Spectral geometry with a cut-off: topological and metric aspects.
\newblock {\em J. Geom. Phys}, 82:18--45, 2014.

\bibitem{dAndrea:2009xr}
F.~D'Andrea and P.~Martinetti.
\newblock A view on optimal transport from noncommutative geometry.
\newblock {\em SIGMA}, 6(057):24 pages, 2010.

\bibitem{Devastato:2013fk}
A.~Devastato, F.~Lizzi, and P.~Martinetti.
\newblock {G}rand {S}ymmetry, {S}pectral {A}ction and the {H}iggs mass.
\newblock {\em JHEP}, 01:042, 2014.

\bibitem{buckley}
A.~Devastato and P.~Martinetti.
\newblock Twisted spectral triple for the standard and spontaneous breaking of
  the grand symmetry.
\newblock {\em arXiv 1411.1320 [hep-th]}.

\bibitem{Englis:2014aa}
M.~Englis, K.~Falk, and B.~Iochum.
\newblock Spectral triples and {T}oeplitz operators.
\newblock {\em arXiv 1402.30610}, 2014.

\bibitem{Martinetti:2011fko}
P.~Martinetti and L.~Tomassini.
\newblock Noncommutative geometry of the {M}oyal plane: translation isometries,
  {C}onnes' distance on coherent states, {P}ythagoras equality.
\newblock {\em Commun. Math. Phys.}, 323(1):107--141, 2013.

\bibitem{rieffel2003}
M.~Rieffel.
\newblock Compact quantum metric spaces.
\newblock {\em Contemporary Mathematics}, 2003.

\bibitem{Walterlivre}
W.~van Suijlekom.
\newblock {\em Noncommutative geometry and particle physics}.
\newblock Springer, 2015.

\end{thebibliography}
